\documentstyle[11pt,moriond,epsfig]{article}

\def\be{\begin{equation}}
\def\ee{\end{equation}}
\def\bea{\begin{eqnarray}}
\def\eea{\end{eqnarray}}

\newcommand{\ecm}{E_{\mathrm cm}}

\newcommand{\fig}{Figure~\ref}
\newcommand{\tab}{Table~\ref}
\newcommand{\as}{$\alpha_s$}
\newcommand{\oas}{$\cal O$($\alpha_s^2$)}
\newcommand{\gev}{\mbox{\,\,Ge\kern-0.2exV}}
\newcommand{\mev}{\mbox{\,\,Me\kern-0.2exV}}

\newcommand{\DW}{Dokshitzer and Webber}
\newcommand{\asb}{$\bar{\alpha}_0$}
\newcommand{\beq}{\begin{equation}}
\newcommand{\eeq}{\end{equation}}
\def\jetset{{\sc Jetset}}

\newcommand{\bmax}{$\left<B_{\mathrm max}\right>$}

\newcommand{\bsum}{$\left<B_{\mathrm sum}\right>$}
\newcommand{\mh}{$\left<M^2_{\mathrm h}/E^2_{\mathrm vis}\right>$}
\newcommand{\mhm}{\left<M^2_{\mathrm h}/E^2_{\mathrm vis}\right>}
\newcommand{\thr}{$\left<1-T\right>$}
\newcommand{\thrm}{\left<1-T\right>}

\begin{document}
\vspace*{4cm}
\title{POWER CORRECTIONS AT LEP}

\author{ D. WICKE }

\address{Fachbereich Physik, Bergische Universit\"at-GH,\\ 
  Gau{\ss}str.~20, 42097 Wuppertal, Germany\\
\tt wicke@cern.ch}

\maketitle\abstracts{
The size of non-perturbative corrections to event shape observables is
predicted to fall like powers of the inverse centre of mass energy.
These power corrections are investigated for different observables from 
$e^+e^-$-annihilation measured at LEP as well as previous experiments. 
The obtained corrections are compared to other approaches and theoretical
predictions. Measurements of \as\ using power corrections are compared to
conventional methods.
}

\section{Introduction}
The process of hadron production in $e^+e^-$-annihilation is usually
depicted by three phases. The first part called the perturbative phase is 
described by perturbative QCD calculations. The second
phase called fragmentation or hadronisation is usually described using Monte
Carlo based models. It is widely hoped that the influence of this phase on
event shape observables can be described by analytical means: Power
corrections. A third phase containing hadron decays is believed to be well
under control. 

Power corrections arise from two different theoretical approaches:
\em Renormalons \em and  \em analytical hadronisation models\em. 
This suggests a connection between the picture of a hadronisation phase and the
theoretical idea of renormalons.

The size of the correction due to the hadronisation process can be seen as a
quality attribute for a specific observables. A small correction implies
that this observable probes the parton structure more directly, allowing a
more precise test of QCD predictions, e.g. measurements of \as. 
As power corrections use few parameters for describing the hadronisation phase,
it is hoped that they lead to a better understanding of its influence.

\section{Simple Power Corrections}
\subsection{Mean Values}
Means of infrared and collinear safe observables can be described by the sum of
the perturbative part and the power correction term
\beq
\left< f \right> = 
\frac{1}{\sigma_{\mathrm tot}}\int f\frac{df}{d\sigma}d\sigma =
\left< f_{\mathrm pert} \right> + \left< f_{\mathrm pow} \right> 
\label{eq_f}
\eeq
where the perturbative prediction in 2nd order can be written as
\beq
\left< f_{\mathrm pert} \right> = A   \frac{\alpha_s(\mu)}{2\pi}+
               \left(A\cdot 2 \pi b_0 \ln\frac{\mu^2}{\ecm^2} + B\right)
                \left(\frac{\alpha_s(\mu)}{2\pi}\right)^2{\rm ,}
\label{eq_fpert_o2}
\eeq
with A and B being given numbers~\cite{NuclPhysB178_412,CERN89-08vol1}, 
$\mu$ being the renormalisation
scale and $b_0=(33-2N_f)/12\pi$.
To investigate the size and type of power corrections,
\begin{eqnarray}
\left < f_{\mathrm pow}\right > & = & \frac{C_1}{\ecm}+ \frac{C_2}{\ecm^2}
\label{eq_pow_simple}
\end{eqnarray}
is used as a simple ansatz.
It is useful to fix $\alpha_s$ in these fits to get comparable power
coefficients.

\begin{figure}
\mbox{\epsfig{file=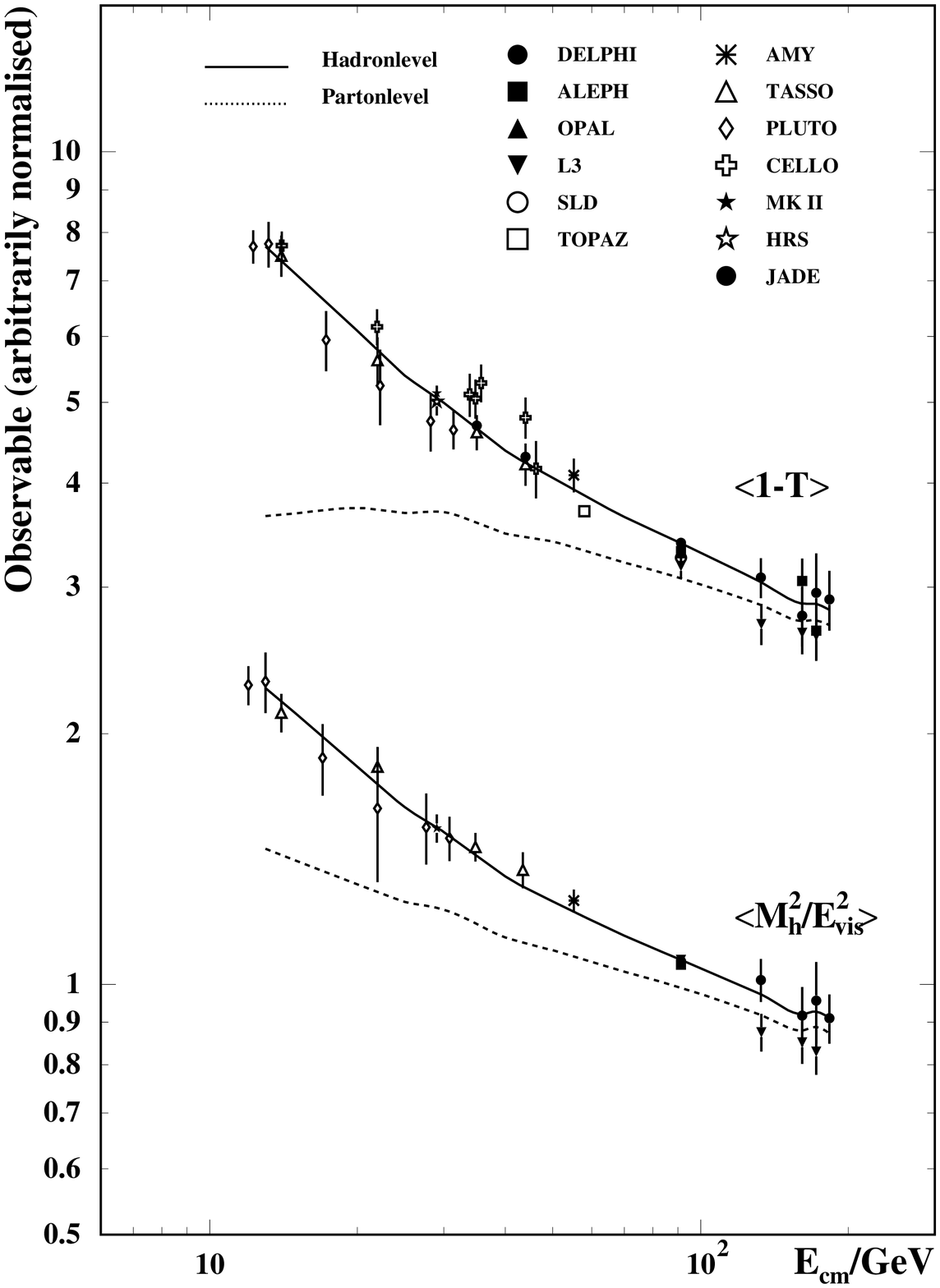,height=11cm}
      \hspace{-1.7cm}\epsfig{file=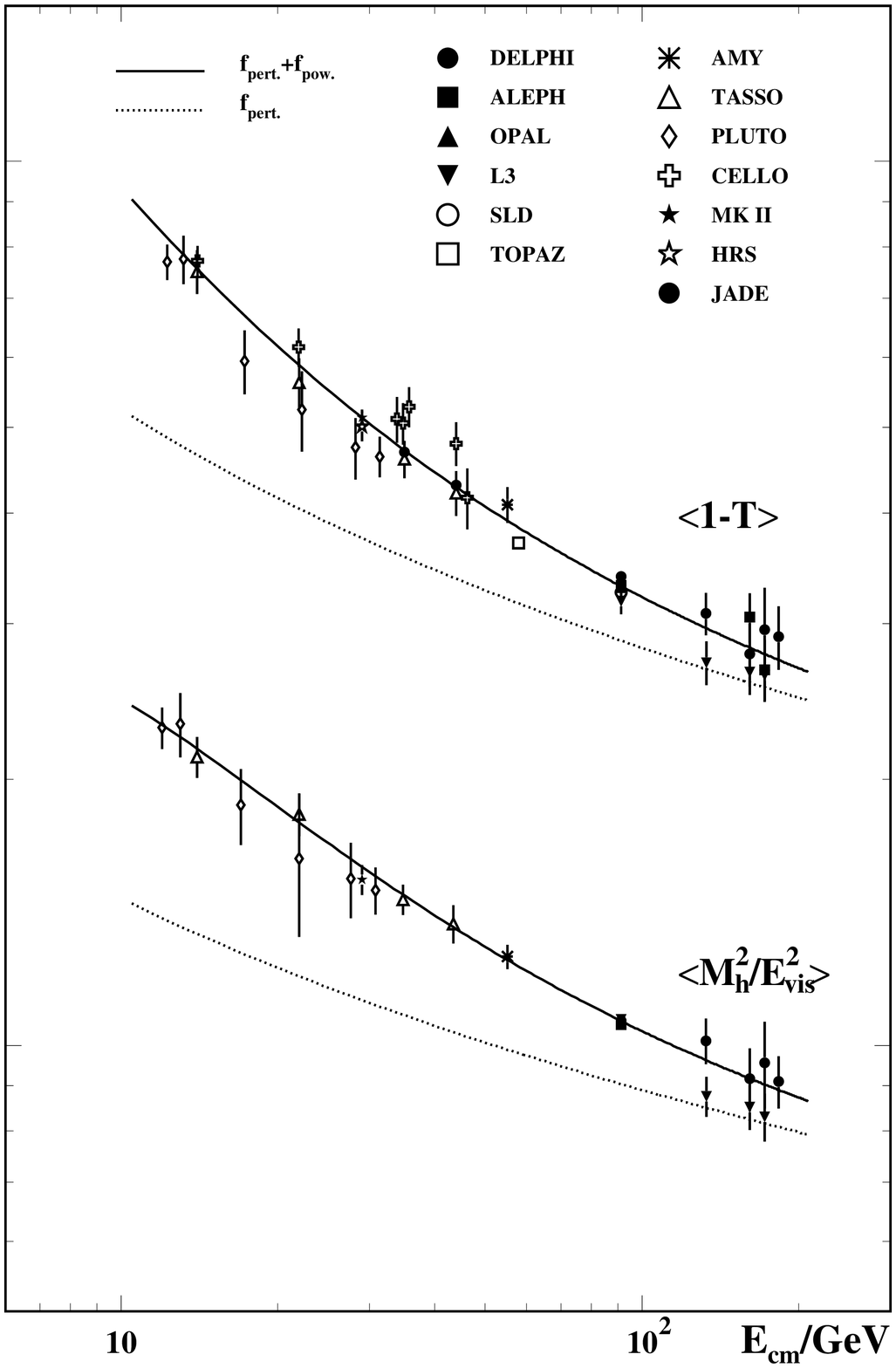,height=11cm}}\\[-1cm]
\mbox{\epsfig{file=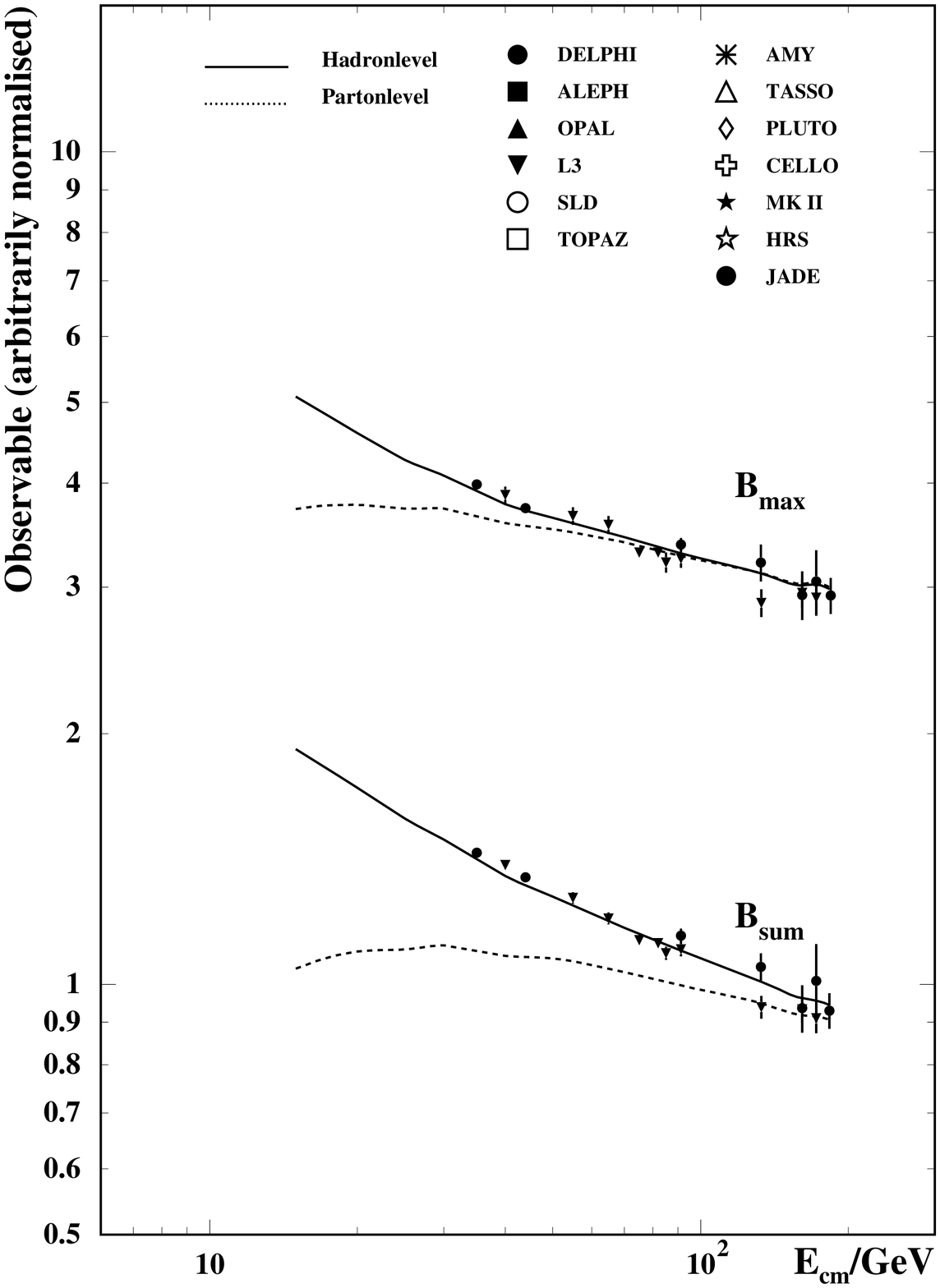,height=11cm}
      \hspace{-1.7cm}\epsfig{file=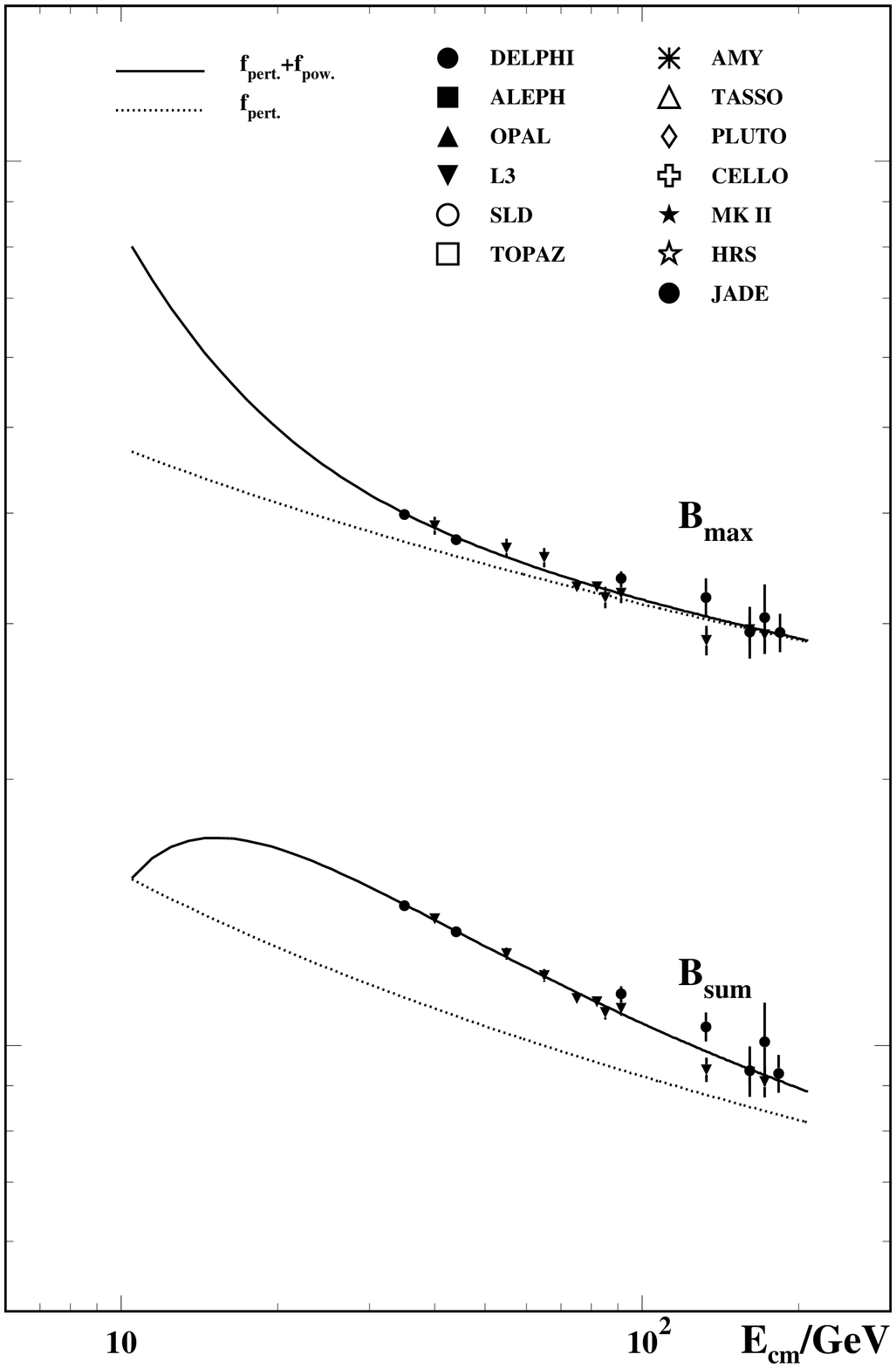,height=11cm}}
\caption{\label{SimpleFig}Measured $\left<1-T\right>$,
$\left<M_h^2/E_{\mathrm vis}^2\right>$, \bsum\ and \bmax\ 
\protect\cite{collection_eventshapes,Acciarri:1997dn,DELPHI97-92conf77,DELPHI98-18conf119}
as a function of the
centre of mass energy. 
On the lefthand-side the solid lines present the hadron level prediction of \jetset 74,
the dotted lines show the parton shower result.
On the righthand-side the solid lines present the results of the fits with
Eqs.~(\ref{eq_f}--\ref{eq_pow_simple}), the dotted lines show the
perturbative part only.
The strong decrease of the \bsum\ prediction at low energies at the right
is caused by the
large and negative $C_2$ coefficient which is poorly determined.
}
\end{figure}

The four mean event shapes in \fig{SimpleFig} 
show qualitative agreement between the
parton levels of the parton shower Monte Carlo and the second order part
resulting from the fit. The corrections of all four means show $1/\ecm$ 
behaviour with only
$1-T$ having large $\chi^2$ due to inconsistent data. Although the $C_2$
coefficients of \thr\ and \bsum\ are not consistent with 0, the
dominant contributions comes from the $1/\ecm$ term. \thr\ and \bsum\
have larger corrections than \mh\ and \bmax. The numbers are given in
\tab{SimpleFig}. 
\begin{table}[b]
\caption{\label{SimpleTab}Coefficients 
from fitting Equations (\ref{eq_f}--\ref{eq_pow_simple})
 to various observables with $\alpha_s(M_Z)=0.118$ fixed.}
\begin{center}
\small
\begin{tabular}{|l|r|r|c||l|r|r|c|}
\hline
Observable &$C_1~ [\gev]$&$C_2~ [\gev^2]$&\rule[-1.2ex]{0pt}{4ex}$\frac{\chi^2}{\mathrm{ndf}}$&Observable &$C_1~ [\gev]$& $C_2~ [\gev^2]$&$\frac{\chi^2}{\mathrm{ndf}}$\\
\hline
\thr&$1.09\pm 0.03$ &$~-3.2\pm 0.9$&$72/38$
&$\left<1\!-\!T\right>_{0.2...0.5}$ & $0.37\pm 0.03$&$0.6\pm 1.2$&$20/9$\\
\mh&$0.67\pm 0.12$ &$~-1.9\pm 0.7$&$21/25$ &&&&\\
\bsum&$1.27\pm 0.09$ &$\!-10.3\pm 4.4$&$21/19$&\raisebox{-0ex}[0ex][0ex]{$\left<\frac{M^2_{\mathrm h}}{E^2_{\mathrm
          vis}}\right>_{0.1...0.5}$}&{ $0.08\pm 0.03$}&{$9.2\pm 0.9$}&{$~8/5$}\\
\bmax & $0.33\pm 0.06$ &$~~0.2\pm 2.3$&$23/20$&&&&\\
\hline
\end{tabular}
\end{center}
\end{table}


\subsection{Cut Integrals and Higher Moments}
To investigate whether the non-perturbative correction of an event depends
on specific values of an observable or not cut integrals 
$\left<x\right>_{a\ldots b}:=\frac{1}{\sigma_{\mathrm tot}}\int_a^b
\frac{d\sigma}{dx}\;dx$
were investigated.
While $\thrm_{0.1\ldots 0.5}$ shows  dominantly $1/\ecm$ 
correction like the full mean,
$\mhm_{0.1\ldots 0.5}$ acquires a $1/\ecm^2$ behaviour.
(\tab{SimpleTab} and \fig{OtherFig}a). 
\begin{figure}
\epsfig{file=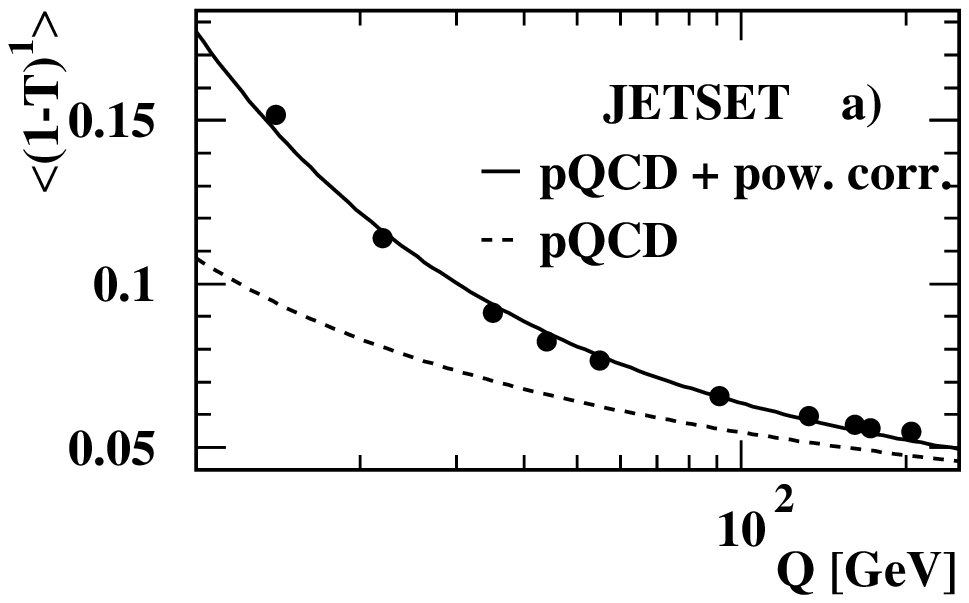,width=5cm}
\epsfig{file=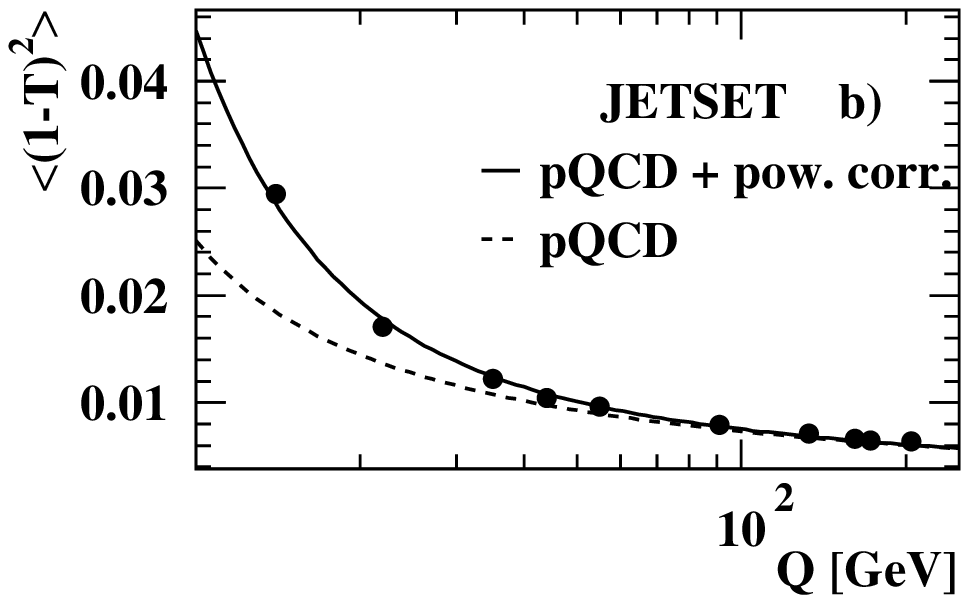,width=5cm}
\epsfig{file=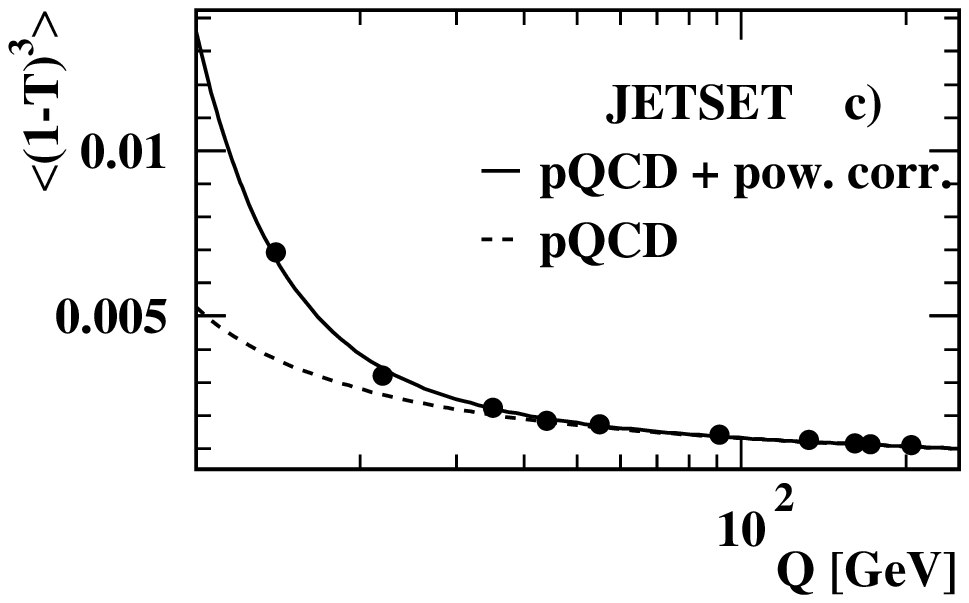,width=5cm}\vspace*{-0.5cm}
\caption{\label{OPALFig}Results of fitting
     Eqs.\ (\ref{eq_f},\ref{eq_fpert_o2},\ref{eq_fpow_opal}) to
     \jetset\ 
     predictions for mean values and higher moments of $1-T$.}
\end{figure}
\begin{table}
\caption{\label{OPALTab}Results of fitting
  Eqs.\ (\ref{eq_f},\ref{eq_fpert_o2},\ref{eq_fpow_opal}) to \jetset\
  predictions for mean values and higher moments of $1-T$.}
\begin{center}
\begin{tabular}{|l|lll|}
\hline
$1-T$& Mean~~~~~~~ & 2nd Moment & 3rd Moment\\
\hline
$\alpha_s(M_Z)$& 0.108 & 0.1121 & 0.1131\\
$\bar\alpha_{n-1}$& 0.380& 1.34 & 5.7   \\
$\chi^2/{\mathrm{ndf}}$& 16&4.8 & 3.0   \\
\hline
\end{tabular}
\end{center}
\end{table}

Another way of emphasising different ranges is to investigate higher moments.
The OPAL
Coll.\ investigated the power dependence of the first three moments of $1-T$
and the $C$-parameter~\cite{OPAL-PN-310}.
\beq
\left < f_{\mathrm pow}\right >  =  a_f \frac{1}{n^2}
\left(\frac{\mu_I}{\ecm}\right)^n
  \left[\bar{\alpha}_{n-1}(\mu_I) - \alpha_s(\mu)
        - \left(b_0 \cdot \ln{\frac{\mu^2}{\mu_I^2}} + 
\frac{K}{2\pi} + \frac{2b_0}{n} \right) \alpha_s^2(\mu) 
\right]\quad{\rm ,}
\label{eq_fpow_opal}
\eeq
was used as power term. 
$\bar{\alpha}_{n-1}$ being a non-perturbative parameter accounting for the
contributions to the event shape below an infrared matching scale $\mu_I$,
$K=(67/18-\pi^2/6)C_A-5N_f/9$ and $a_f=4C_f/\pi$.
Beside \as\ these formulae contain $\bar{\alpha}_{n-1}$ 
 as the only free parameters.
The results shown for \thr\ in \fig{OPALFig} and Table \ref{OPALTab}  
indicate that the assumed
power law of $1/\ecm^n$ for the $n$-th moment does work. 

\section{Other Approaches}
\begin{figure}[p]
\mbox{
\unitlength1mm
\scriptsize
\begin{picture}(0,0)
\put(13,0){a)}
\put(63,0){b)}
\put(113,0){c)}
\put(63,37){$\chi^2/{\mathrm ndf}=$}
\put(73,33){$186/35$}
\put(63,12){$\chi^2/{\mathrm ndf}=74/20$}
\put(115,18){$\chi^2/{\mathrm ndf}=$}
\put(125,14){$75/35$}
\put(115,37){$\chi^2/{\mathrm ndf}=$}
\put(125,33){$70/20$}
\end{picture}
\epsfig{file=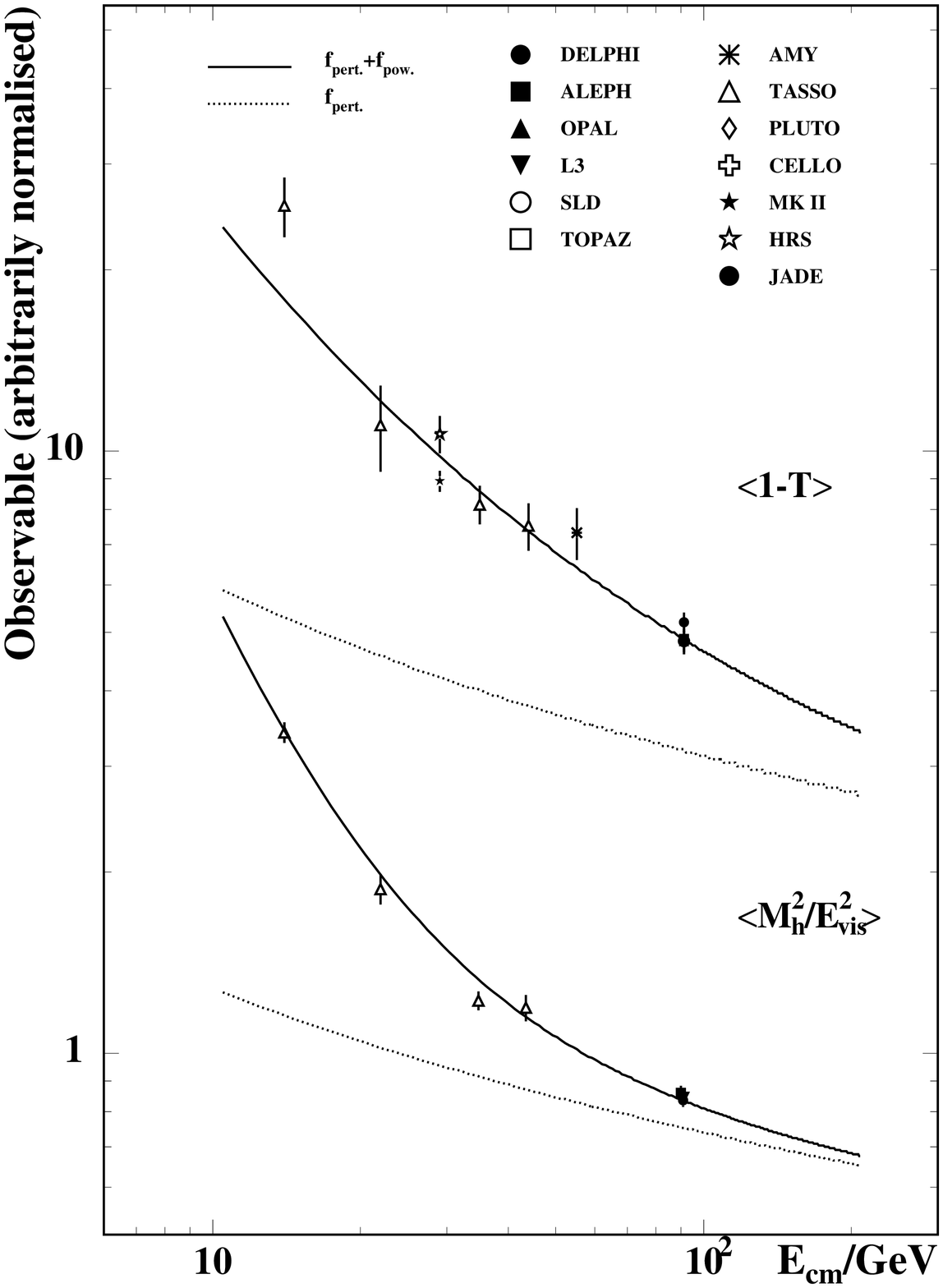,width=5cm}
\epsfig{file=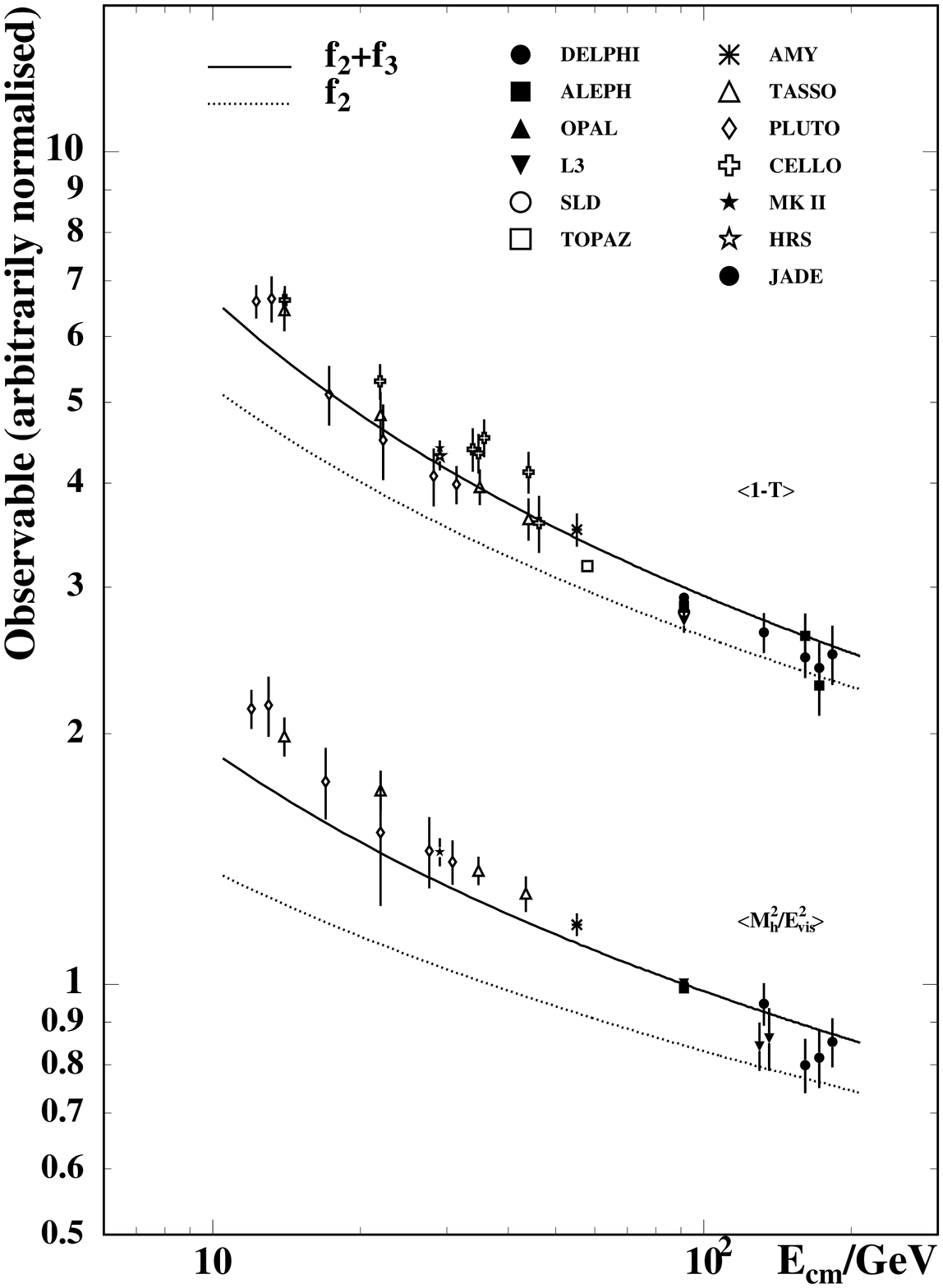,width=5cm}
\epsfig{file=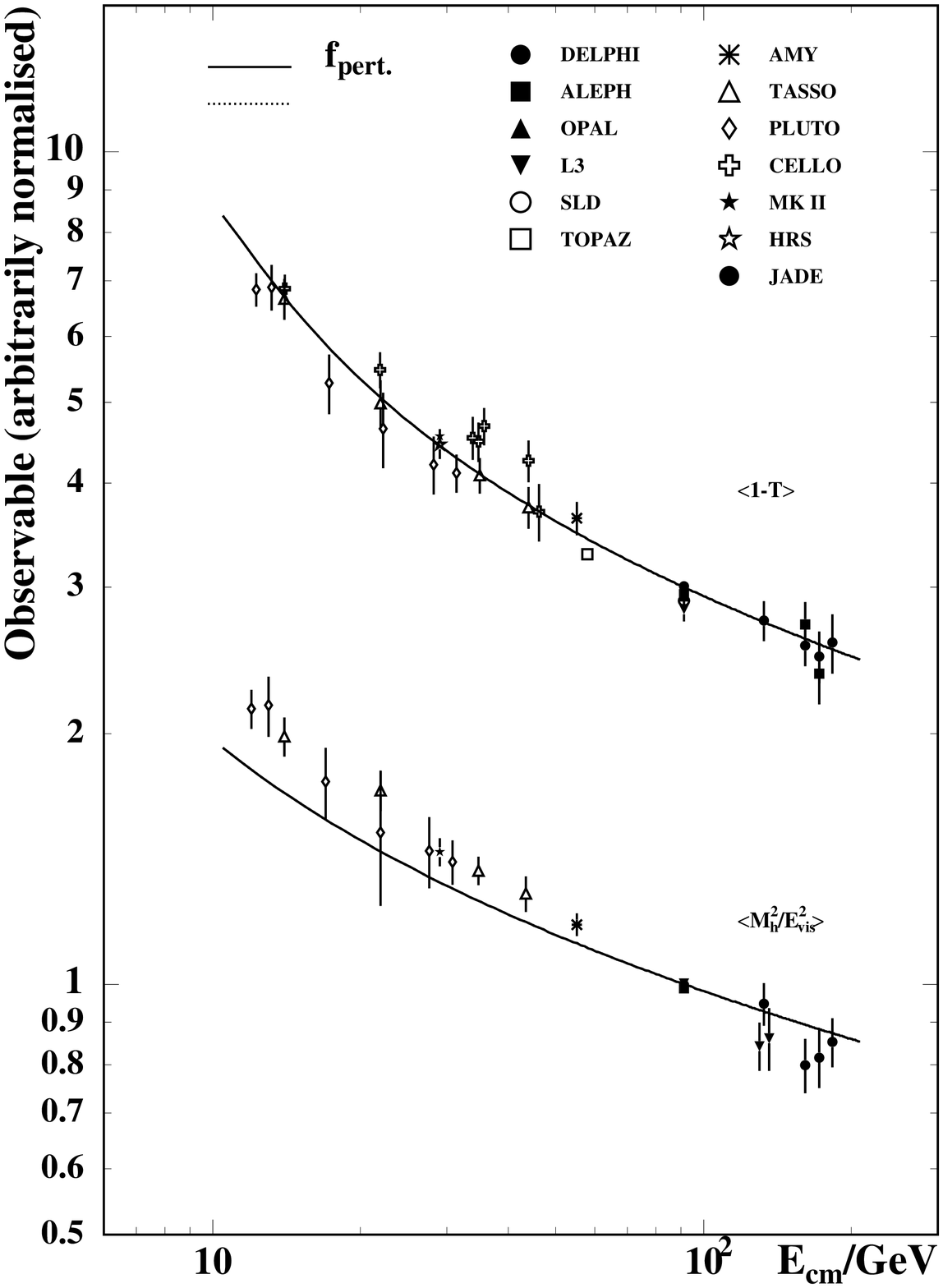,width=5cm}}
\caption{\label{OtherFig}Fit results for cut variables (left), fitted 3rd
  order coefficient (middle) and fitted
  renormalisation scale (right).
  The continuous line represents the complete prediction. In the left plot the
  dashed line gives the perturbative contribution. In the middle plot it
  represents the 2nd order contribution to the full 3rd order curve.
}
\end{figure}

To strengthen the need for power corrections in the description of data, a few
other approaches were investigated:
Fitting only the perturbative prediction including a 3rd order coefficient as
free parameter yields very large $\chi^2$ and a 3rd order coefficient 
$C={\cal O}(1000)$ for \thr\ and \mh. Thus a 3rd order calculation does not
give an improved prediction on the energy dependence neglecting hadronisation
effects. (\fig{OtherFig}b). 

Leaving the renormalisation scale $x_\mu=\mu^2/\ecm^2$ free, shows
ambivalent results. While for \thr\ the fit is reasonable and
leads to a scale $x_\mu$ that is consistent with the one obtained from
fits to distributions, the fit for \mh\ cannot describe the data.
(\fig{OtherFig}c). 

To obtain the functional type of power corrections needed, one can also fit
with an arbitrary power $p$: 
\begin{equation}
\left < f_{\mathrm pow}\right >  =  \frac{C_p}{\ecm^p}
\end{equation}
For \thr\ one gets $p=0.98\pm0.19$~\cite{ALEPH-LP97-258} in perfect agreement
with the previous results.

\section{Determination of \as\ using Power Corrections}

The analytical  power ansatz for non-perturbative corrections by 
\DW~\cite{PhysLettB352_451,hep-ph/9510283}
is used by DELPHI to determine \as\ from mean event
shapes~\cite{DELPHI97-92conf77,DELPHI98-18conf119}. 
The power correction of this ansatz is given by
\beq
\left < f_{\mathrm pow}\right >  =  a_f \cdot
\frac{\mu_I}{\ecm}
  \left[\bar{\alpha}_0(\mu_I) - \alpha_s(\mu)
        - \left(b_0 \cdot \ln{\frac{\mu^2}{\mu_I^2}} + 
\frac{K}{2\pi} + 2b_0 \right) \alpha_s^2(\mu) 
\right]\quad{\rm ,}
\label{eq_fpow_dw}
\eeq
with $\mu_I$, $K$, $a_f$ and $b_0$ as given above.
In order to measure \as\ from individual high energy data the free parameter
\asb\ has to be known.
\begin{table}[p]
\caption{\label{tab_mess_fit}Determination of \asb. For $\ecm\ge M_Z$
only DELPHI measurements are included in the fit.
The first error is the statistical
error from the fit, the second one is the scale error.}
\begin{center}
\begin{tabular}{|c|c|c|c|c|} \hline
Observable    & $\bar{\alpha}_0$         
                &$\alpha_s(M_Z)$ 
                        & $\Lambda_{\overline{MS}}$[\mev]\rule[-1ex]{0pt}{3ex}
                & $\chi^2/\mathrm{ndf}$ \\ \hline
$\left<1-T\right>$ 
& $0.531 \pm 0.012 \pm 0.003$
                & $0.1187 \pm 0.0017\pm 0.0060$
                        & $235 \pm 22 \pm 70$ & 42/23 \\
$\left<M_h^2/E_{\mathrm vis}^2\right>$ 
 & $0.434 \pm 0.010 \pm 0.010$
                & $0.1144 \pm 0.0012\pm 0.0041$
                        & $184 \pm 13 \pm 40$ & 4.0/14 \\ \hline 
\end{tabular}
\end{center}
\end{table}

To infer \asb\ a combined fit of \as\ and \asb\ to a large set of
measurements at different energies~\cite{collection_eventshapes}
is performed. For $\ecm\ge M_Z$ only DELPHI measurements are included in
the fit.
The resulting values of \asb\ are summarised in \tab{tab_mess_fit}.
%
The extracted \asb\ values are around 0.5 as expected in~\cite{hep-ph/9510283}.
The numerical values are, however, incompatible with each other. 
So the assumed universality~\cite{hep-ph/9510283} is not valid
to the precision that is accessible from the data. 
Therefore \asb\ is determined for $\left<1-T\right>$ and
$\left<M_h^2/E_{\mathrm vis}^2\right>$ individually. The scale error is
obtained from varying the renormalisation scale factor $x_\mu=\mu^2/\ecm^2$ 
from 0.25 to 4 and the infrared matching scale from 
$1 \gev$ to $3 \gev$.

\begin{figure}[p]
\mbox{\epsfig{file=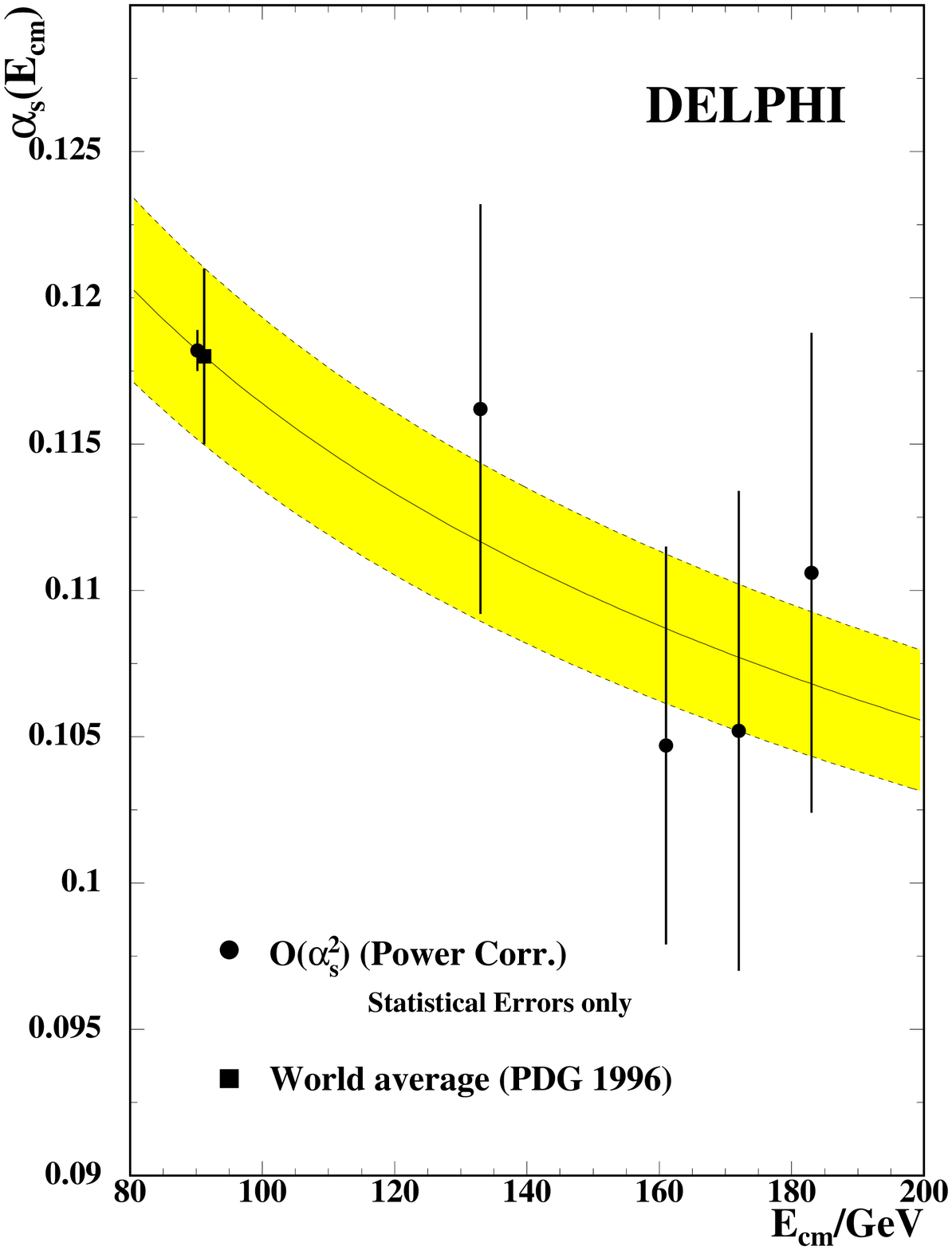,width=4.6cm}
\hspace{-1.06cm}\epsfig{file=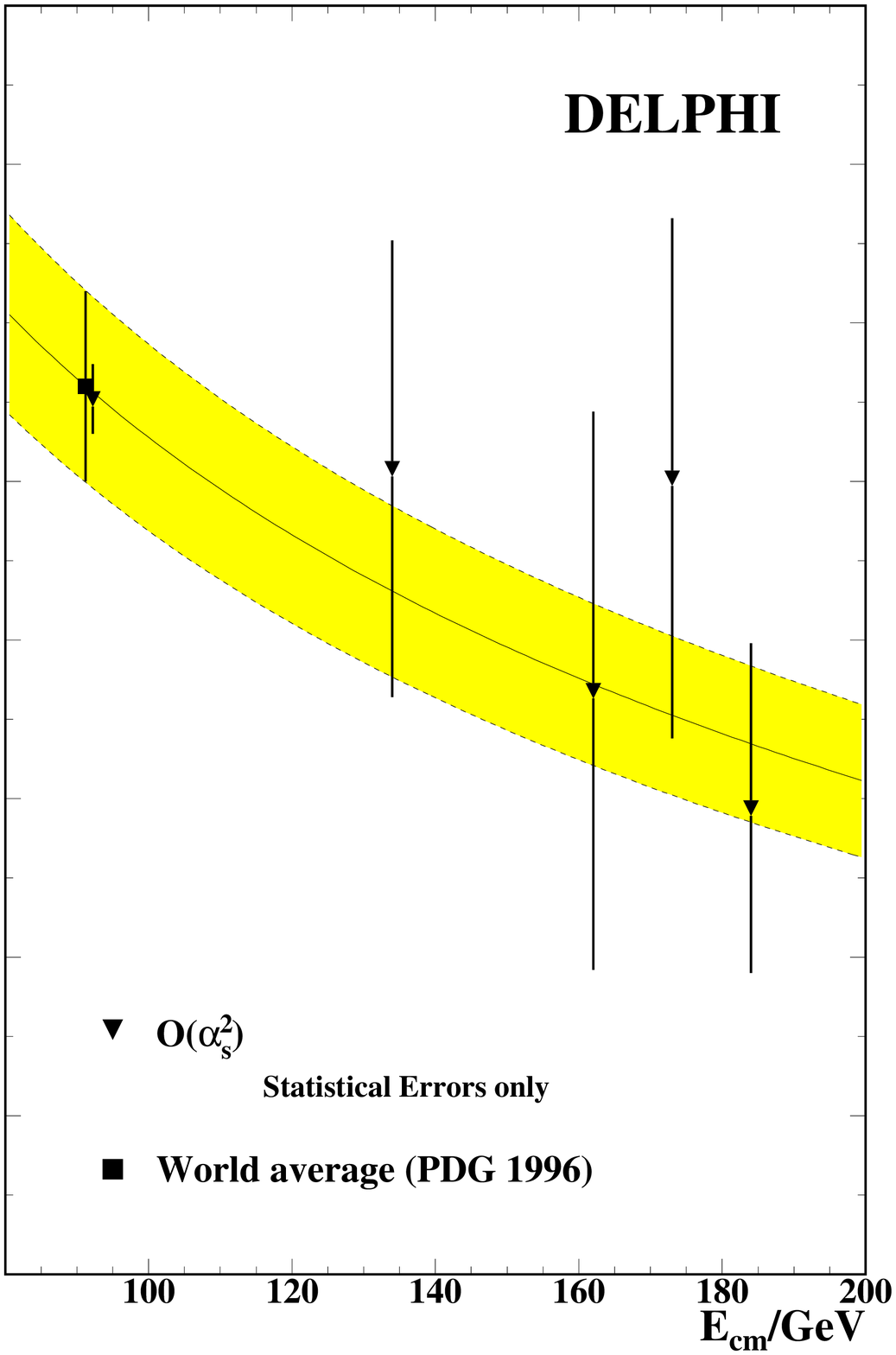,width=4.6cm}
\hspace{-1.06cm}\epsfig{file=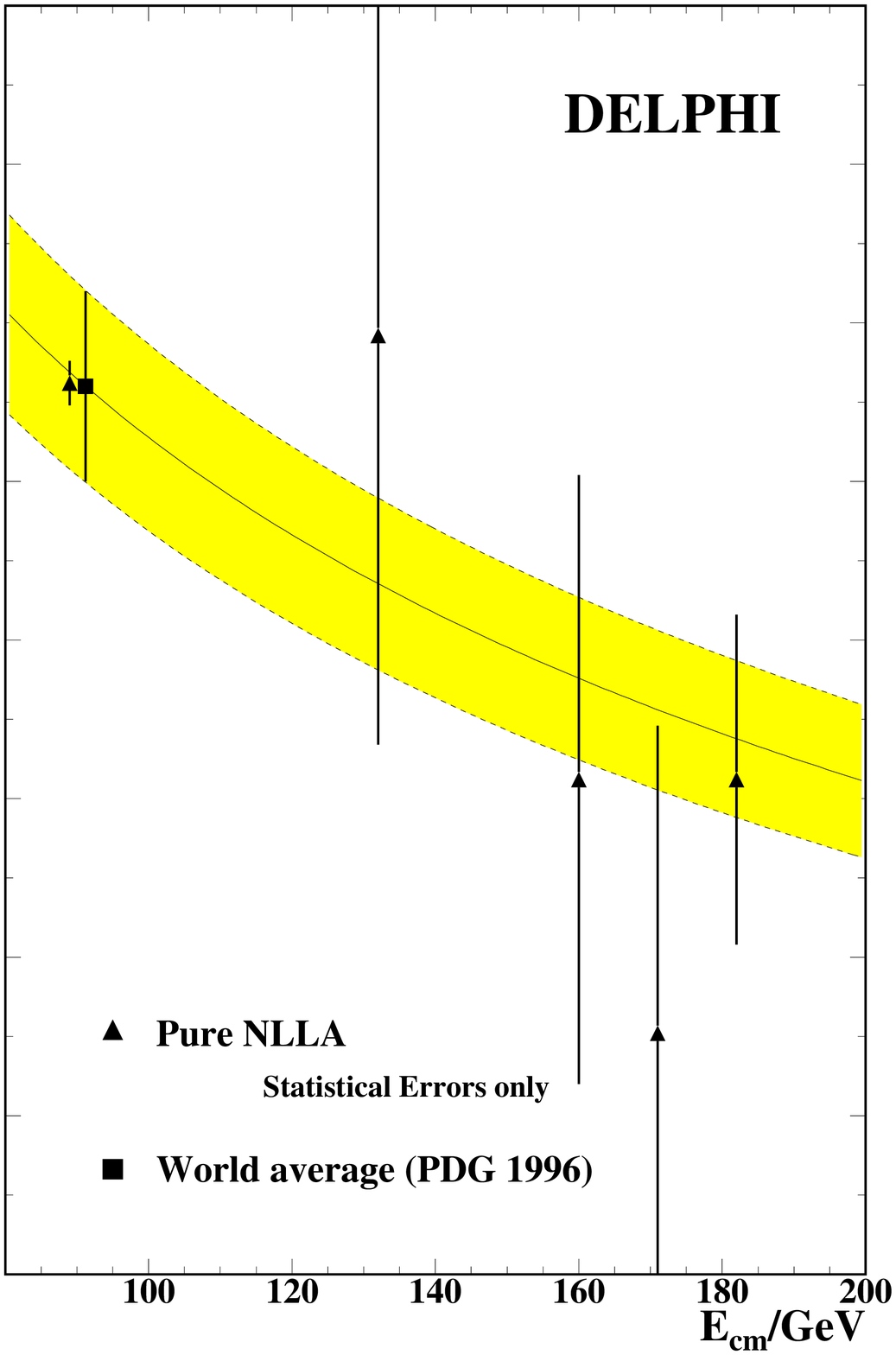,width=4.6cm}
\hspace{-1.06cm}\epsfig{file=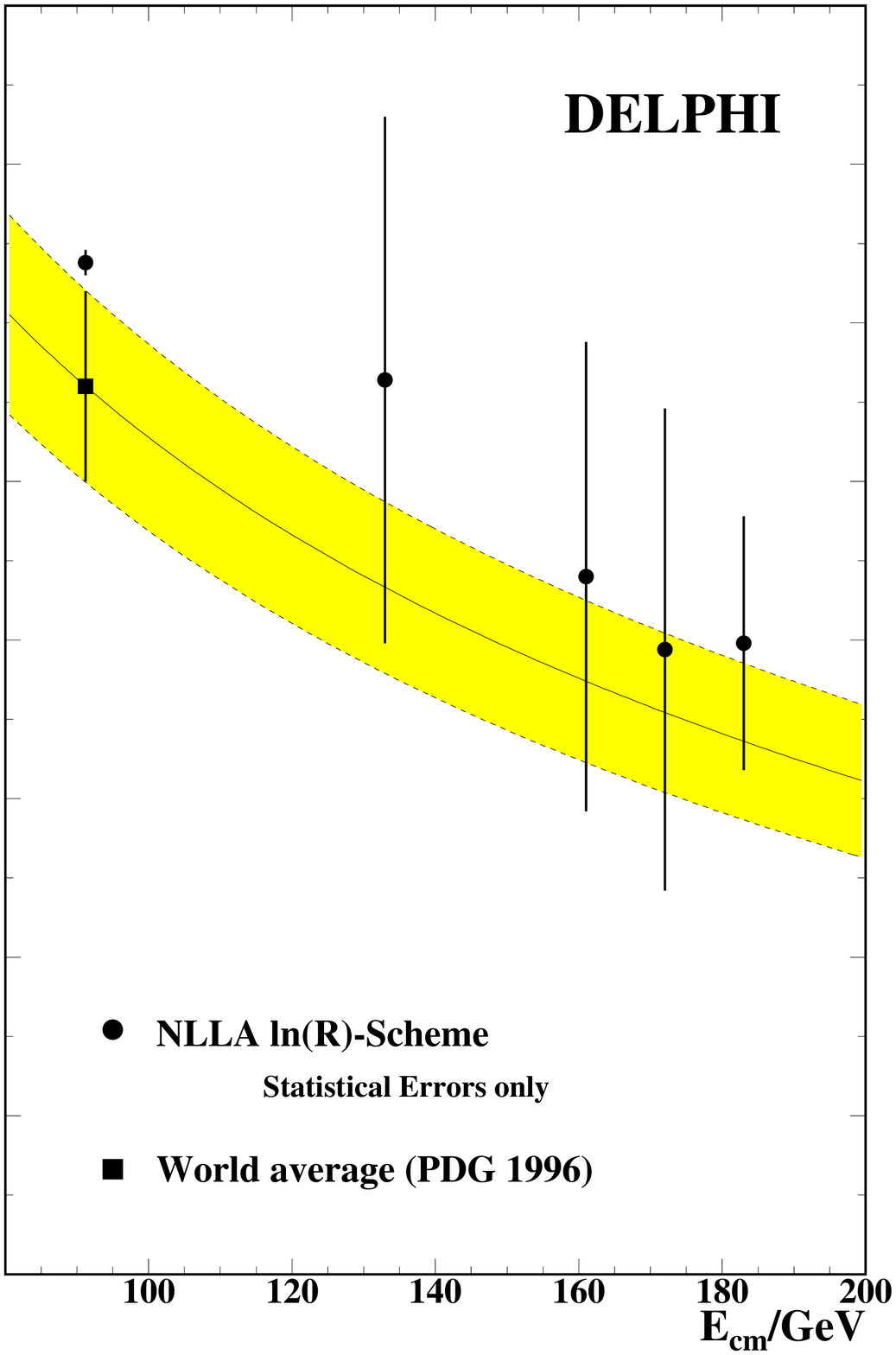,width=4.6cm}}
\caption{\label{fig_dasdE}Energy dependence of \as\ as obtained from 
                     mean event shapes (left) compared to 
                     \as\ obtained from distributions.
                     The errors shown are statistical
                     only. The band shows the QCD
                     expectation of extrapolating the world average to
                     other energies.}
\end{figure}

After having fixed \asb, the \as\ values corresponding to the high
energy data points can be calculated from 
Eqs.~(\ref{eq_f},\ref{eq_fpert_o2},\ref{eq_fpow_dw}). 
\as\ is calculated for both
observables individually and then combined with an unweighted average.
The resulting \as\ values and the QCD expectation 
are shown in the leftmost plot of
\fig{fig_dasdE}.

\begin{table}[p]
\caption{\label{tab_dasdE}Running of $\alpha_s$ ($d\alpha/dE_{\mathrm{cm}}$)
                     between 91\gev\ and 183\gev.}
\begin{center}
\begin{tabular}{|l|l|l|l|l|}
\hline
{{Means plus}} & \multicolumn{3}{|c|}{Distributions plus}&\\ 
{{Power corr.}} & \multicolumn{3}{|c|}{Monte Carlo based hadronisation  models}&QCD pred.\\  
{{\oas}}        & \oas &  NLLA  & \oas+NLLA&  \\
\hline
{{$-0.13 \pm 0.05$}} & $-0.13 \pm 0.05$ & $-0.15 \pm 0.05$ & $-0.13 \pm 0.04$&$-0.13\quad\cdot 10^{-3}\gev^{-1}$\\
\hline
\end{tabular}
\end{center}
\end{table}

The \as\ values follow the QCD expectation.
Fitting a straight line to the energy dependence results in a slope,
which agrees very well with the QCD expectation of a running
\as\ between 91\gev\ and 183\gev.
In \fig{fig_dasdE} and \tab{tab_dasdE} this result 
is further-on compared to \as\ measurements obtained from distributions using 
Monte Carlo based hadronisation models. The comparison shows that the running
measured from means using power correction and the one obtained from standard
methods give consistent results and comparable errors.

\section{Prospects}
Many of the recent developments in the field of power corrections are not yet
included in the presented experimental works. Beside the calculation of Milan
factors, which will not influence the
consistency of the experimental 
results much, 
some of the predicted power coefficients $a_f$ were corrected in the
last months~\cite{Moriond98LucentiSalam}. 
First tests show that the consistency of measured \asb\ and \as\
values improve with these new predictions.

Thus the experimental fits have to be repeated and with increasing theoretical
understanding consistent experimental results hopefully will arise.


\section*{References}
\newcommand{\esdcollection}
{   
 ALEPH  Coll. {\em Phys. Lett.} {\bf B284} (1992) 163;
 ALEPH Coll. {\em Z. Phys.} {\bf C55} (1992) 209;
 AMY  Coll. {\em Phys. Rev. Lett.} {\bf 62} (1989) 1713;
 AMY Coll. {\em Phys. Rev.} {\bf D41} (1990) 2675;
 CELLO Coll. {\em Z. Phys.} {\bf C44} (1989) 63;
 HRS Coll. {\em Phys. Rev.} {\bf D31} (1985) 1;
 JADE Coll. {\em Z. Phys.} {\bf C25} (1984) 231;
 JADE Coll. {\em Z. Phys.} {\bf C33} (1986) 23;
 P.A. Movilla Fernandez, et. al. and the JADE Coll. 
 {\em Eur. Phys. J.} {\bf C1} (1998) 461;
 L3 Coll. {\em Z. Phys.} {\bf C55} (1992) 39;
 Mark II Coll. {\em Phys. Rev.} {\bf D37} (1988) 1; 
 Mark II Coll. {\em Z. Phys.} {\bf C43} (1989) 325;
 MARK J Coll. {\em Phys. Rev. Lett.} {\bf 43} (1979) 831;
 OPAL Coll. {\em  Z. Phys.} {\bf C59} (1993) 1;
 PLUTO Coll. {\em Z. Phys.} {\bf C12} (1982) 297; 
 SLD Coll. {\em Phys. Rev.} {\bf D51} (1995) 962;
 TASSO Coll. {\em Phys. Lett.} {\bf B214} (1988) 293;
 TASSO Coll. {\em Z. Phys.} {\bf C45} (1989) 11;
 TASSO Coll. {\em Z. Phys.} {\bf C47} (1990) 187; 
 TOPAZ Coll.  {\em Phys. Lett.} {\bf B227} (1989) 495;
 TOPAZ Coll. {\em Phys. Lett.} {\bf B278} (1992) 506; 
 TOPAZ Coll. {\em Phys. Lett.} {\bf B313} (1993) 475}

\bibliography{QCD} 
\end{document}